# Resolving Transition Metal Chemical Space: Feature Selection for Machine Learning and Structure-Property Relationships

Jon Paul Janet[1] and Heather J. Kulik[1,*]

[1]*Department of Chemical Engineering, Massachusetts Institute of Technology, Cambridge, MA 02139*

ABSTRACT: Machine learning (ML) of quantum mechanical properties shows promise for accelerating chemical discovery. For transition metal chemistry where accurate calculations are computationally costly and available training data sets are small, the molecular representation becomes a critical ingredient in ML model predictive accuracy. We introduce a series of revised autocorrelation functions (RACs) that encode relationships between the heuristic atomic properties (e.g., size, connectivity, and electronegativity) on a molecular graph. We alter the starting point, scope, and nature of the quantities evaluated in standard ACs to make these RACs amenable to inorganic chemistry. On an organic molecule set, we first demonstrate superior standard AC performance to other presently-available topological descriptors for ML model training, with mean unsigned errors (MUEs) for atomization energies on set-aside test molecules as low as 6 kcal/mol. For inorganic chemistry, our RACs yield 1 kcal/mol ML MUEs on set-aside test molecules in spin-state splitting in comparison to 15-20x higher errors from feature sets that encode whole-molecule structural information. Systematic feature selection methods including univariate filtering, recursive feature elimination, and direct optimization (e.g., random forest and LASSO) are compared. Random-forest- or LASSO-selected subsets 4-5x smaller than RAC-155 produce sub- to 1-kcal/mol spin-splitting MUEs, with good transferability to metal-ligand bond length prediction (0.004-5 Å MUE) and redox potential on a smaller data set (0.2-0.3 eV MUE). Evaluation of feature selection results across property sets reveals the relative importance of local, electronic descriptors (e.g., electronegativity, atomic number) in spin-splitting and distal, steric effects in redox potential and bond lengths.

# 1. Introduction

Computational high-throughput screening is key in chemical and materials discovery[1-11], but high computational cost has limited chemical space exploration to a small fraction of feasible compounds[12-13]. Machine-learning (ML) models have emerged as alternative approaches especially for efficient evaluation of new candidate materials[14] or potential energy surface fitting and exploration through sophisticated force field models[15-22]. Examples of recent ML applications in computational chemistry include exchange-correlation functional development[23-24], general solutions to the Schrödinger equation[25], orbital free density functional theory[26-27], many body expansions[28], acceleration of dynamics[29-31], band-gap prediction[32-33], and molecular[34-35] or heterogeneous catalyst[36] and materials[37-40] discovery, to name a few.

Essential challenges for ML models to augment or replace first-principles screening are model selection and transferable feature set identification. For modest sized data sets, descriptor set selection is especially critical[41-43] for successful ML modeling. Good feature sets should[42] be cheap to compute, as low dimensional as possible, and preserve target similarity (i.e. materials with similar should properties have similar feature representations). Within organic chemistry, structural descriptors such as a Coulomb matrix[44] or local descriptions of the chemical environment and bonding[43, 45] have been useful to enable predictions of energetics as long as a relatively narrow range of elements (e.g., C, H, N, O, F) is considered. These observations are consistent with previous successes in evaluating molecular similarity[46], force field development[47], quantitative structure-activity relationships[48], and group additivity[49] theories on organic molecules.

Descriptors that work well for organic molecules have proven unsuitable for inorganic materials[50] or molecules[51]. This lack of transferability can be readily rationalized: it is well-known[51-54] that some electronic properties of transition metal complexes (e.g., spin state splitting) are much more sensitive to direct ligand atom identity that dominates ligand field strength[55-56]. Unlike organic molecules, few force fields have been established that can capture the full range of inorganic chemical bonding[57]. The spin-state- and coordination-environment-dependence of bonding[58] produces a higher-dimensional space that must be captured by sophisticated descriptors or functions. In spite of these challenges, suitable ML models for inorganic chemistry will be crucial in the efficient discovery of new functional materials[59-60], for solar energy[61], and for catalyst discovery[62].

With the unique challenges of inorganic chemistry in mind, we recently trained a neural network to predict transition metal complex quantum mechanical properties[51]. From several candidate descriptor sets, we demonstrated good performance, i.e., 3 kcal/mol root mean squared error for spin-splitting and 0.02-0.03 Å for metal-ligand bond lengths, of heuristic, topological-only near-sighted descriptors. These descriptors required no precise three-dimensional information and outperformed established organic chemistry ML descriptors that encode more whole-complex information.

In this work, we introduce systematic, adaptable-resolution heuristic and topological descriptors that can be tuned to encode molecular characteristics ranging from local to global. As these descriptors require no structural information, rapid ML model prediction without prior first-principles calculation is possible, and such ML models can improve structure generation[63] through bond length prediction[51, 64]. We apply this adaptable descriptor set to both organic and inorganic test sets, demonstrating excellent transferability. We use rigorous feature selection tools to quantitatively identify optimal locality and composition in machine learning feature sets for predicting electronic (i.e., spin-state and redox potential) and geometric (i.e., bond length) properties. The outline of the rest of this work is as follows. In Sec. 2, we review our new descriptors, methods for subset selection, and the ML models trained in this work. In Sec. 3, we provide the Computational Details of first-principles data sets and associated simulation methodology. In Sec. 4, we present Results and Discussion on the trained ML models for spin-state splitting, bond-lengths, and ionization/redox potentials. Finally, in Sec. 5, we provide our Conclusions.

## 2. Approach to Feature Construction and Selection

### 2a. Autocorrelation Functions as Descriptors.

Autocorrelation functions[65] (ACs) are a class of molecular descriptors that have been used in quantitative structure-activity relationships for organic chemistry and drug design[66-68]. ACs are defined in terms of the molecular graph, with vertices for atoms and unweighted (i.e., no bond length or order information) edges for bonds. Standard ACs[65] are defined as:

$$P_d = \sum_i \sum_j P_i P_j \delta(d_{ij}, d) \qquad (1)$$

where $P_d$ is the AC for property $P$ at depth $d$, $\delta$ is the Dirac delta function, and $d_{ij}$ is the bond-wise path distance between atoms $i$ and $j$. Alternatives to the eqn. 1 AC sums are motivated and discussed in Sec. 2b. The AC depth $d$ thus encodes relationships between properties of atoms separated by $d$ bonds; it is zero if $d$ is larger than the longest molecular path, and 0-depth ACs are just sums over squared properties. The five atomic, heuristic properties used in our ACs are: i) nuclear charge, $Z$, as is used in Coulomb matrices[69]; ii) Pauling electronegativity, $\chi$, motivated by our previous work[64]; iii) topology, $T$, which is the atom's coordination number; iv) identity, $I$, that is 1 for any atom, as suggested in Ref. [12]; and v) covalent atomic radius, $S$. Although i, ii, and v are expected to be interrelated, the $S$ quantity uniquely imparts knowledge of spatial extent, and covalent radii follow different trends than $Z$ or $\chi$ (e.g. the covalent radius of Co is larger than Fe and Ni).

ACs are compact descriptors, with $d+1$ dimensions per physical property encoded at maximum depth $d$, that depend only on connectivity and do not require Cartesian or internal coordinate information. Although inclusion of geometric information improves predictive capabilities of machine learning models in organic chemistry[70], reliance on structural information requires explicit calculation or knowledge of it prior to ML prediction, which is not practical for transition metal complexes. AC sets also are vectorial descriptors that are invariant with respect to system size and composition, unlike frequently-used symmetry functions[19], bag-of-bonds[71], and Coulomb matrices[69, 72].

Despite their promise in therapeutic drug design[66-68] or in revealing inorganic complex structure-property relationships[64], ACs have not yet been tested as features in machine learning models that predict quantum mechanical properties. We first apply ACs to the QM9 database[73] of 134k organic (C, H, O, N, and F elements) molecules consisting of up to nine heavy atoms. This database contains B3LYP/6-31G-calculated properties, including atomization energies and HOMO-LUMO gaps, making it a frequent test set for machine learning models and descriptor sets[43, 70, 74]. The QM9 test set allows us to both identify if there is an optimal maximum depth for ACs and to determine the baseline predictive capability of ACs in comparison to established descriptors[69, 71-72]. Throughout this work, we score feature sets by training Kernel ridge regression (KRR) models[75] with a Gaussian kernel. KRR is a widely-employed[69, 71-72] ML model, that has produced sub-kcal/mol out-of-sample property prediction error on large organic databases and crystals[76-77]. We have selected KRR for the i) ease of retraining, ii) transparency of differences in KRR models[75], as predictions are related to arrangement of data points in feature space, and iii) wide use of KRR in computational chemistry[69, 71-72, 76-77] (Supporting Information Text S1).

First, we test the effect of increasing the maximum AC depth to incorporate increasingly nonlocal ACs on AE prediction test set errors using a 1,000 molecule training set repeated five times (Figure 1). We evaluate prediction test set mean unsigned error (MUE) on the remaining 133k molecules in the QM9 set. Test set MUEs first decrease with increasing depth from 18 kcal/mol MUE at zero-depth (i.e., only sums of constituent atom properties) and reach a minimum of 8.8 kcal/mol MUE at maximum three-depth ACs. Without any further feature reduction, maximum three-depth ACs ($3d$-ACs) correspond to a 20-dimensional feature set (i.e., 4 length scales x 5 properties). Increasing the maximum depth beyond three increases test errors slightly up to 9.2 kcal/mol for maximum six-depth ACs (Figure 1). Minimum train/test MUEs with $3d$-ACs emphasizes the length scale of chemically relevant effects, in line with previous observations[51-52, 70], and increasing train/test MUEs due to the addition of more poorly correlating non-local descriptors emphasizes the importance of careful feature selection (Sec. 2c). Regardless of maximum depth chosen, AC-derived prediction accuracy is impressive since the KRR model is trained with < 1% of the QM9 data set, which has a large overall AE mean absolute deviation of 188 kcal/mol.

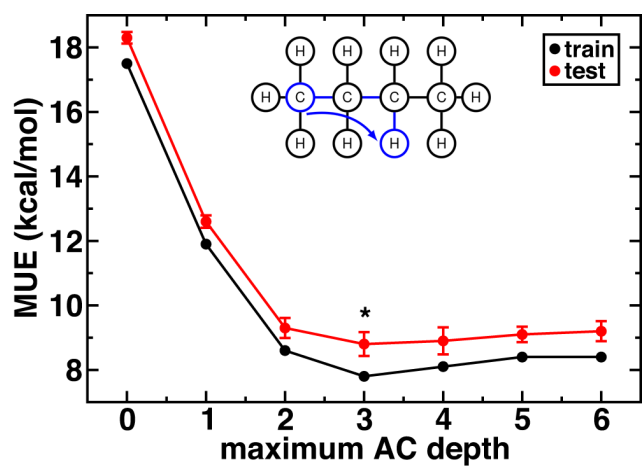

**Figure 1.** Train (black line) and test (red line) MUEs (in kcal/mol) for QM9[73] AEs predicted by KRR models trained on AC feature sets with increasing maximum depth. Each model is trained on 1,000 molecules and tested on the 133k remaining molecules. Error bars on test set error correspond to standard deviations from training the KRR model on five different samples, and the red circles correspond to the mean test error. The lowest MUE maximum-depth, 3, is indicated with an asterisk. An example of a term in a 3-depth AC is shown on butane in inset.

We now compare 3$d$-AC performance and learning rates (i.e., over increasingly large training sets) to i) the Coulomb matrix eigenspectrum[69] (CM-ES) representation, which is an easy to implement 3D-derived descriptor[72]; ii) the recently-developed[70] $2^B$ descriptor that, like ACs, does not require explicit 3D information and encodes connectivity and bond order information for atom pairs; and iii) and more complex[70] $12^{NP}3^B4^B$ descriptors. The $12^{NP}3^B4^B$ descriptors, which encode a continuous, normal distribution of bond distances for each bond-type in a system-size invariant manner, require 3D information but have demonstrated the best to-date performance for QM9 AEs.[70] We trained the CM-ES KRR model using the recommended[72] Laplacian kernel, but we selected a Gaussian kernel for 3$d$-ACs after confirming it produced lower MUEs (Supporting Information Text S2). We compare these models to reported performance of $2^B$ and $12^{NP}3^B4^B$ descriptors from the literature.[70]

For the largest 16,000 molecule training set, the 3$d$-AC test set MUEs are 68% and 43% lower than CM-ES and $2^B$ descriptors, respectively. The 3$d$-AC descriptors are only outperformed by $12^{NP}3^B4^B$ by 74% or 4.5 kcal/mol, owing to the bond distance information encoded in this set (Figure 2 and Supporting Information Table S1). This improved performance of $12^{NP}3^B4^B$ and other comparably-performing[70] descriptors (e.g., superposition of atomic densities[45] or the many-body tensor representation[76]) comes at a severe cost of requiring accurate geometries before predictions can be made, whereas 3$d$-AC significantly outperforms the previous best-in-class topology-only descriptors set $2^B$. Learning rates (i.e., training-set size test set MUE dependence) are comparable among 3$d$-AC, $2^B$, and $12^{NP}3^B4^B$ descriptors but slightly steeper for the poorer performing CM-ES representation (Figure 2). For dipole moment prediction, 3$d$-AC performs nearly as well as $12^{NP}3^B4^B$: the 3$d$-AC test MUE at 1,000 training points is only 2% higher than $12^{NP}3^B4^B$ and 19% higher at 16,000 training points (Supporting Information Table S2). Thus, ACs are promising size-invariant, connectivity-only descriptors for machine learning of molecular properties. However, we have previously observed limited transferability of organic representations for inorganic complexes[51], and we next identify the transferability of our present descriptors as well as beneficial inorganic chemistry adaptations.

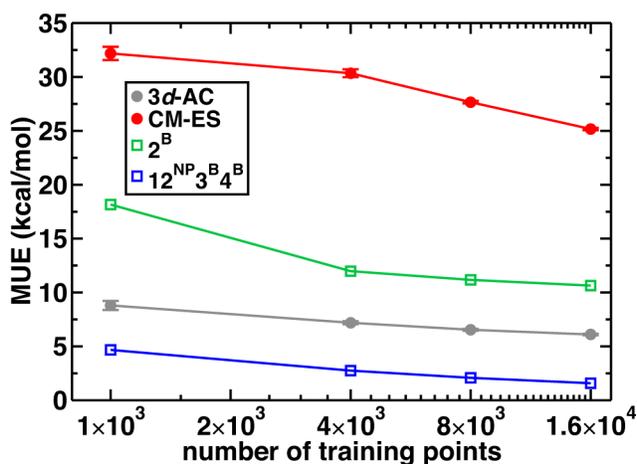

**Figure 2.** Training set size dependence of test set MUEs (in kcal/mol) for KRR model prediction of QM9[73] AEs for four feature sets. In all cases, the test set consists of the remainder of the 134k molecule set not in the training set. For the maximum 3-depth autocorrelation (3$d$-AC, gray circles) and Coulomb matrix eigenspectrum[69] (CM-ES, red circles) trained in this work, standard deviations (error bars) and mean test errors are reported from training results on five samples selected for each training set size. The $2^B$ (green open square) and $12^{NP}3^B4^B$ (blue open square) KRR test set MUEs from literature[70] are provided for comparison.

**2b. Revised Autocorrelations for Transition Metal Complexes.**

We previously proposed[51] a mixed continuous (e.g., electronegativity differences) and discrete (e.g., metal and connecting atom identity) set of empirical, topological descriptors (referred to in this work as MCDL-25) that emphasized metal-proximal properties for predictive modeling of transition metal complexes with an artificial neural network. The MCDL-25 set is metal-focused in nature with the longest range effects only up to two bonds through a truncated Kier shape index[78]. This imparted good accuracy (i.e., root mean squared error, RMSE, of 3 kcal/mol) for spin-state splitting predictions and superior transferability to test set molecules with respect to commonly-employed descriptors[69] used in machine learning for organic chemistry that encode complete, 3D information.

In addition to standard ACs (eqn. 1 in Sec. 2a), we now introduce revised ACs (RACs) inspired by descriptors in the metal-focused MCDL-25 set. In these RACs, we both restrict where the sums in eqn. 1 *start* (i.e., to account for potentially greater importance of the metal and inner coordination sphere) and which other atoms are in the *scope* (Figure 3). In the extended notation of the broader AC set, the standard ACs *starts* on the full molecule (*f*) and has all atoms in the *scope* (*all*), i.e., $_{all}^{f}P_d$. As in ref. [64], we compute restricted-*scope* ACs that separately evaluate axial or equatorial ligand properties:

$$_{ax/eq}^{f}P_d = \frac{1}{|\text{ax/eq ligands}|} \sum_{i}^{n_{ax/eq}} \sum_{j}^{n_{ax/eq}} P_i P_j \delta(d_{ij}, d) \qquad (2)$$

where $n_{ax/eq}$ is the number of atoms in the corresponding axial or equatorial ligand and properties are averaged within the ligand subtype. We introduce restricted-*scope*, metal-centered (*mc*) descriptors, in which one of the atoms, *i*, in the *i,j* pair is a metal center:

$$_{all}^{mc}P_d = \sum_{i}^{mc} \sum_{j}^{all} P_i P_j \delta(d_{ij}, d) \qquad (3)$$

For the complexes in this work there is only one metal-center, which simplifies the sum, but there is no inherent restriction to a single metal center (see green arrows in Figure 3).

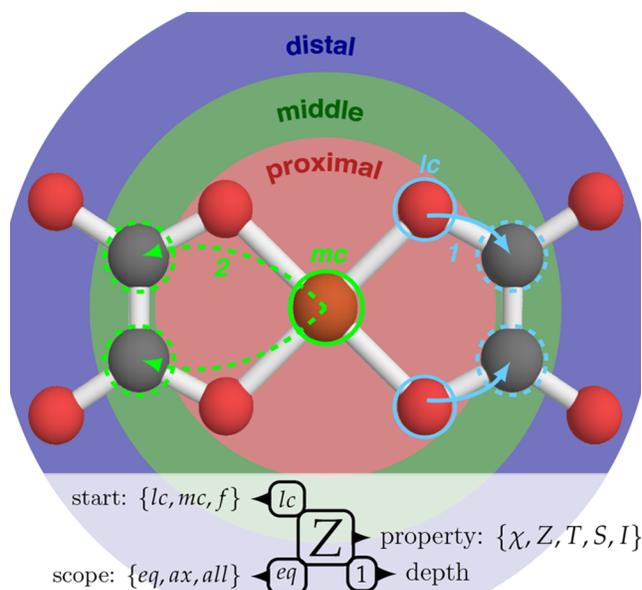

**Figure 3.** Schematic of ACs in the equatorial plane of an iron octahedral complex with two eq. oxalate ligands shown in ball and stick representation (iron is brown, oxygen is red, and carbon is gray). Regions of the molecule used to classify descriptors are designated as proximal (metal and first coordination shell, in red), middle (second coordination shell, in green) and distal (third shell and beyond, in blue) throughout the text. Light green circles and arrows depict terms in a 2-depth *mc* RAC (e.g., $^{mc}_{eq}Z_2$), and the light blue circles and arrows depict terms in a 1-depth *lc* RAC (e.g., $^{lc}_{ax}Z_1$).

A second restricted-*scope*, metal-proximal AC definition is the ligand-centered (*lc*) sum in which one of the atoms, *i*, in the *i,j* pair is the metal-coordinating atom of the ligand:

$$^{lc}_{ax/eq}P_d = \frac{1}{|ax/eq\ ligands|} \frac{1}{|lc|} \sum_i^{lc} \sum_j^{n_{ax/eq}} P_i P_j \delta(d_{ij}, d) \qquad (4)$$

We average the ACs over all *lc* atoms and over all ligands in order to treat ligands of differing denticity on equal footing (see light blue arrows in Figure 3).

Inspired by our previous success[51, 79] in employing electronegativity differences between atoms to predict electronic properties, we also modify the AC definition, *P'*, to property differences rather than products for a minimum depth, *d*, of 1:

$$_{ax/eq/all}^{lc/mc}P'_d = \sum_i^{lc\,or\,mc} \sum_j^{scope} (P_i - P_j)\delta(d_{ij}, d) \tag{5}$$

where *scope* can be axial, equatorial, or all ligands, the *start* must be *lc* or *mc* because a sum of differences over *all* will be zero, and these ACs are not symmetric so the ordering of indices *i,j* is enforced for consistency.

We combine all six types of AC or RAC start/scope definitions (*f/all*; *mc/all*; *lc/ax*; *lc/eq*, *f/ax*; and *f/eq*, eqns. 1-5) with both products and differences of the five atomic properties for depths from zero, where applicable, to maximum depth *d*. There are 6*d*+6 descriptors for six product AC/RACs (eqns. 1-4) with each of the five atomic properties (i.e., a total of 30*d*+30 product AC/RACs). For difference RACs (eqn. 5), there are no zero-depth descriptors, and three non-trivial start/scope definitions (*mc/all*; *lc/ax*; and *lc/eq*), producing 3*d* descriptors for all of the atomic properties excluding *I*, giving *12d* difference descriptors for a total of 42*d*+30 product or difference RACs. These ACs represent a continuous vector space that is increasingly nonlocal with increased maximum *d* and dimension invariant with respect to system size. This descriptor set also does not depend on any 3D information, which is valuable for structure prediction[51, 64].

We classify relative locality of ACs into three categories (see Figure 3): 1) proximal: depends only on atom types and connectivity in first coordination shell; 2) middle: depends on information from two coordination shells; and 3) distal: all remaining descriptors based on the molecular graph. This broad AC set naturally recovers well-known quantities: i) $_{all}^{mc}I_1$ is the metal coordination number and ii) $_{all}^{f}I_0$ is the total number of atoms. We also recover continuous descriptor analogues to the variables in MCDL-25[51]: i) $_{all}^{mc}Z_0$ is the metal identity, ii) $_{ax/eq}^{lc}Z_0$ is the coordinating atom identity, and iii) $_{ax/eq}^{lc}\chi'_1$ is $\Delta\chi$. Some ACs are redundant (e.g., $_{all}^{mc}I_1$ and $_{all}^{mc}T_0$ are the same). Before model training, all ACs are normalized to have zero-mean and unit variance based on training data, and any constant features in training data are filtered out.

## 2c. Feature Selection Methods.

Feature reduction from a large descriptor space improves the ratio of training points to the dimension of the feature space, decreasing training time and complexity[80] for non-linear models (e.g., neural networks) or improving predictions in kernel-based methods with isotopic kernels by eliminating uninformative features. In linear models, feature reduction increases stability, transferability, and out-of-sample performance[80]. Reducing feature space, without impact on model performance[80], is also useful[81] for providing insight into which characteristics are most important for determining materials properties. Starting from $n$ observations (e.g., spin-state splitting, bond length, or redox potential) of $y_{data}(x_i)$ and molecular descriptors $x_i \in \mathbb{R}^m$ in an $m$-dimensional feature space, $\mathcal{X}_m$, we use established[81] feature selection techniques to obtain a lower-dimensional representation of the data, $\mathcal{X}_d \in \mathcal{X}_m$, that maximizes out-of-sample model performance while having the smallest possible dimension.

Feature selection techniques may be broadly classified[81] as (Figure 4): 1) simple filters, 2) wrapper methods, and 3) direct optimization or shrinkage methods[75]. Type 1 univariate filtering (UVF) acts on each descriptor individually, discarding those that fail a statistical test (here, the p-value for a linear model being above a cutoff of 0.05). UVF is amenable to very high-dimensional problems[81] but neglects interactions between descriptors that may occur[80], and the significance test in a linear model may not relate well to the final machine learning model.

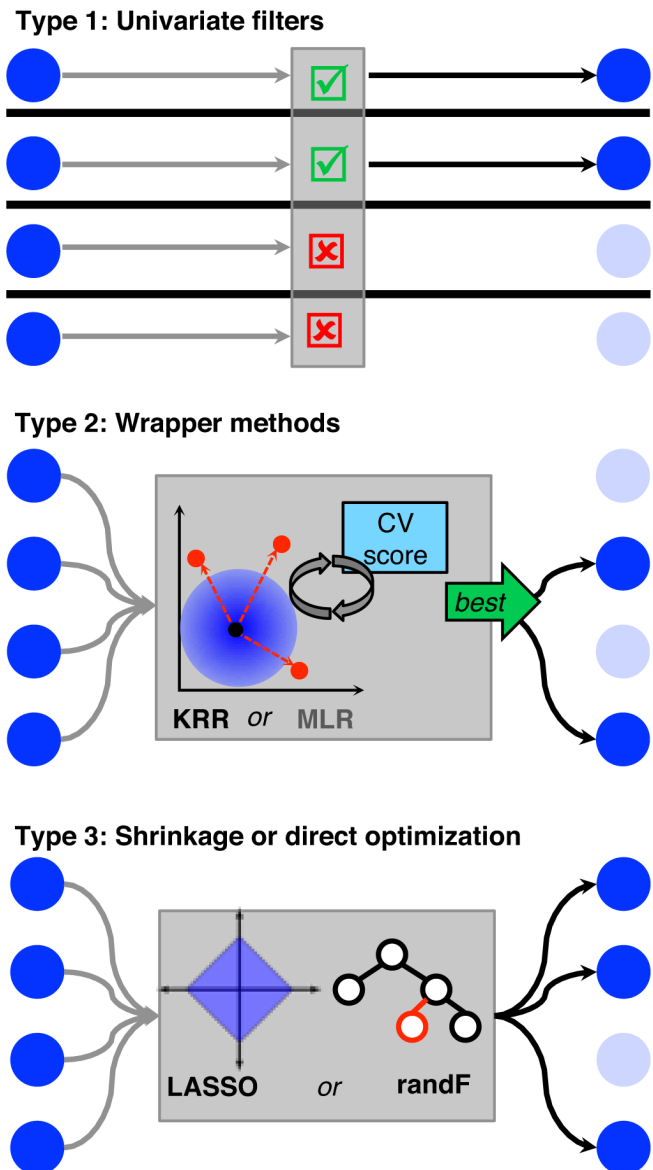

**Figure 4.** Schematic of three main types of feature selection approaches with retained and input features represented by dark blue circles. Type 1 (top) univariate filters evaluate features one at a time; type 2 (middle) wrapper methods train a model (e.g., KRR or MLR) and use a cross validation score to recursively eliminate features; and type 3 (bottom) shrinkage or direct optimization models such as LASSO and random forests (randF) carry out one-shot feature selection and regularization or model training, respectively.

Type 2 wrapper methods require multiple steps[80-81]: iterative feature subset choice along with model training and scoring (Figure 4). Combinatorial testing of every possible subset is only feasible for small feature sets (e.g., < 40 variables with simple predictive models[75]). The model used in training and scoring is flexible, but the repeated model training time may become prohibitive. Stepwise search[82], with greedy recursive feature addition or elimination (i.e., RFA or RFE) on most improvement or least penalty, respectively, or randomized searches less prone to local minima[81], are employed for larger feature sets. Cross-validation (CV) scoring, which is unaffected by feature space size changes, will usually produce a minimum for an optimal number of features[80]. We recently used[64] RFE with an embedded linear model to select variables to use in multiple linear regression (MLR) to identify four key RACs from a larger 28-dimensional space for redox potential prediction. In this work, we primarily employ RFE-MLR to select features to be used for KRR training, despite potentially eroded model transferability between MLR and KRR. The fine hyperparameter grid search needed to produce a robust KRR model at each RFE iteration would take around 30 days in parallel on a 4-core Intel 3.70 GHz Core i7-4820K when starting from a large (ca. 150) descriptor set, making some initial reduction in feature space necessary for practical RFE-KRR (Supporting Information Text S3).

Type 3 shrinkage or direct optimization methods use regularization (e.g., elastic net or L1-regularized linear regression, LASSO[83]) or a model (e.g. random forests) that determines variable importance in one shot during training, making Type 3 methods much more computationally efficient than Type 2. However, it remains uncertain if the typically lower complexity of the combined feature-selection and fitting model (e.g, L1 regularized regression in LASSO) produces results that are transferable to the subsequent ML model to be trained (e.g., KRR). In this work, we use an elastic net, a generalization of LASSO that we previously used to select descriptors for machine learning models[51], in which a blend of L2 and L1 regularization is applied[84], giving the loss function as:

$$L(W) = \|xW - y_{data}(x)\|_2^2 - \lambda\left(\alpha\|W\|_1 + (1-\alpha)\|W\|_2^2\right) \quad (6)$$

Here, $W$ are the regression coefficients, $\lambda$ is the regularization strength, and $\alpha \in [0,1]$ interpolates between ridge ($\alpha=0$) and LASSO ($\alpha=1$) regression. Higher $\alpha$ aggressively reduces the feature space, and the best $\alpha$ is selected by cross-validation with $\lambda$, with intermediate $\alpha$ often favored for balancing prediction with feature reduction[75].

Random forests[85], which are based on an ensemble of sequential binary decision trees, are another Type 3 method (Figure 4). Each tree is trained on a bootstrapped data sample and uses a random input variable set. Integrated feature selection is achieved by comparing tree performance when descriptors are randomly permuted[86] to yield an importance score for each descriptor and discard those below a threshold value. Here, we use 1,000 trees and discard descriptors with an increase of < 1% (or higher, where specified) in normalized MSE on out-of-model samples upon removal (see convergence details in Supporting Information Figures S1-4).

We now compare feature selection methods on our transition metal complex data sets, as judged by performance on 60%-40% and 80%-20% train-test partitions for the larger *spin-splitting* and smaller *redox* data set (see Sec. 3a), respectively. Feature selection is only carried out on the training data. All analysis is conducted in R version 3.2.3[87]. We use the kernlab[88] package for KR regression, CVST[89] for cross-validation, glmnet[90] for elastic net regression, caret[91] for feature selection wrapper functions and randomForest[92] for random forests. All kernel hyperparameter values are provided in Supporting Information Tables S3-S6.

## 3. Computational Details

### 3a. Organization of data sets.

Feature selection and model training is carried out on two data sets of single-site octahedral transition metal complexes, which were generated from extension of data collected in previous work[51, 64] (Figure 5). For both sets, the complexes contain $Cr^{2+/3+}$, $Mn^{2+/3+}$, $Fe^{2+/3+}$, $Co^{2+/3+}$, or $Ni^{2+}$ first row transition metals. High-spin (H) and low-spin (L) multiplicities were selected for each metal from those of the isolated ion in National Institute of Standards and Technology atomic spectra database[93]: triplet-singlet for $Ni^{2+}$, quartet-doublet for $Co^{2+}$ and $Cr^{3+}$, quintet-singlet for $Fe^{2+}$ and $Co^{3+}$, quintet-triplet for $Cr^{2+}$ and $Mn^{3+}$ (due to the fact that there is no data available for $Mn^{3+}$ singlets[93]), and sextet-doublet for $Mn^{2+}$ and $Fe^{3+}$. For all data sets, the molSimplify[63] code was used to generate starting geometries from the above metals and a ligand list (ligands provided in Supporting Information Table S7). Incompatible ligand combinations are disabled (e.g., equatorial porphyrin ligands can occur once and only with monodentate axial ligands).

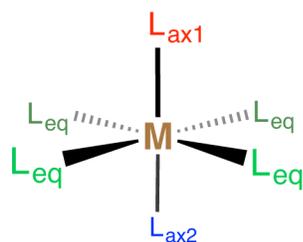

|  | spin-split (1345) | redox (226) | |
|---|---|---|---|
|  |  | new (185) | Fe-N (41) |
| M | Cr,Mn,Fe,Co,Ni | Cr,Mn,Fe,Co | Fe |
| L types | 16 | 5 | 41 |
| L CA | C,N,O,S,Cl | C,N,O | N |
| L denticity | 1 to 4 | 1 | 1 to 2 |
| symmetry | ax≠eq | ax1≠ax2≠eq | ax=eq |
| properties | $\Delta E_{H-L}$, min($R_L$) | $\Delta E_{III-II}$, $E^0$ | |

**Figure 5.** (top) Schematic of octahedral transition metal complex illustrating possible unique ligands (one equatorial ligand type, $L_{eq}$, and up to two axial ligand types, $L_{ax1}$ and $L_{ax2}$) in the *spin-splitting* and *redox* data sets. (bottom) Characteristics of each data set: metal identity, number of ligand types (L types), connecting atom identity of the ligand to the metal (L CA), range of denticities (L denticity), ligand symmetry corresponding to the schematic complex representation, and associated quantum mechanical properties. *Spin-splitting* and *redox* Fe-N sets were previously published[51, 64], but the "new" subset of the *redox* data set was generated in this work.

The *spin-state splitting* data set[51] consists of 1345 unique homoleptic or heteroleptic complexes with up to one unique axial and equatorial ligand type with ligands selected from 16 common ligands of variable ligand field strength, connecting atom identity, and denticity (Figure 5). For this data set, the structures were evaluated using hybrid density functional theory (DFT) at 7 percentages of Hartee-Fock (HF) exchange from 0 to 30% in 5% increments. This set was previously used to train models that predict i) the adiabatic, electronic spin-state splitting energy, $\Delta E_{H-L}$, ii) the exchange sensitivity of the spin-state splitting, and iii) the spin-state dependent minimum metal-ligand bond lengths (e.g., $min(R_L)$ or $min(R_H)$) that differ from the average metal-ligand bond length only for distorted homoleptics or heteroleptic complexes. In this work, we only train and test models on $\Delta E_{H-L}$ and $min(R_L)$.

The *redox* data set (226 unique structures) is comprised of 41 previously studied[64] Fe-nitrogen monodentate and bidentate homoleptic complexes and 185 newly generated structures (Figure 5 and Supporting Information Table S8). The new complexes were obtained by generating combinations of metals (Cr, Mn, Fe, Co) and five small, neutral monodentate ligands (CO, pyridine, water, furan, and methyl isocyanate) with up to two axial ligand types and one equatorial ligand type. Axial ligand disengagement occurred during optimization in several of the 300 theoretically possible cases, reducing the final data set (Supporting Information). In all cases, we calculate the M(II/III) redox couple starting from the adiabatic ionization energy of the reduced complex's ground state spin:

$$\Delta E_{III-II} = E_{III} - E_{II} \qquad (7)$$

At minimum, this ionization energy requires M(II) low-spin and high-spin geometry optimizations as well as the selected lowest energy M(III) state that differs by a single electron detachment (Supporting Information Table S9).

To compute the redox potential, we also include solvent and thermodynamic (i.e. vibrational enthalpy and zero point vibrational energy) corrections in a widely adopted thermodynamic cycle approach[94-96]. We estimate the M(II/III) redox potential in aqueous solution at 300 K, $\Delta G_{solv}$:

$$\Delta G_{solv} = G_{gas}(M(III)) - G_{gas}(M(II)) + \Delta G_s(M(III)) - \Delta G_s(M(II)) \quad (8)$$

where $G_{gas}$ is the gas phase energy with thermodynamic corrections and $\Delta G_s$ is the solvation free energy of the gas phase structure. We then compute the redox potential:

$$E^0 = -\frac{\Delta G_{solv}}{nF} \quad (9)$$

where the number of electrons transferred is $n=1$ and $F$ is Faraday's constant.

### 3b. First-principles Simulation Methodology.

Our simulation methodology was the same for all generated data sets. All DFT calculations employ the B3LYP hybrid functional[97-99] with 20% HF exchange ($a_{HF} = 0.20$), except for cases where HF exchange is varied[52] while holding the semi-local DFT exchange ratio constant. We use the LANL2DZ effective core potential[100] for all transition metals, bromine, and iodine and the 6-31G* basis for the remaining atoms. The use of a modest basis set is motivated by our previous observations[64] that extended basis sets did not substantially alter trends in redox or spin-state properties. Gas phase geometry optimizations were conducted using the L-BFGS algorithm implemented in the DL-FIND[101] (for the *spin-splitting* data set) or in translation rotation internal coordinates[102] (for the *redox* data set) interfaces to TeraChem[103-104] to the default tolerances of 4.5x10$^{-4}$ hartree/bohr for the maximum gradient and 1x10$^{-6}$ hartree for the change in self-consistent field (SCF) energy between steps. All calculations were spin-unrestricted with virtual and open-shell orbitals level-shifted[105] by 1.0 eV and 0.1 eV, respectively, to aid SCF convergence to an unrestricted solution. Deviations of <S²> from the expected value by more than 1 μ_B led to exclusion of that data point from our data set. The aqueous solvent environment, where applicable, was modeled using an implicit polarizable continuum model (PCM) with the conductor-like solvation model (COSMO[106-107]) and ε=78.39. The solute cavity was built using Bondi's van der Waals radii[108] for available elements and 2.05 Å for iron, both scaled by a factor of 1.2. Vibrational entropy and zero-point corrections were calculated from a numerical Hessian obtained with built-in routines in TeraChem[103-104].

## 4. Results and Discussion

### 4a. Spin Splitting Energy.

We evaluate our RACs (i.e., both standard ACs and the modified *start*, *scope*, and difference ACs defined in Sec. 2b) for KRR training on the *spin-splitting* data set and compare to both previous MCDL-25 descriptors[51] and widely-employed[69, 72] Coulomb-matrix-derived descriptors. Based on our results for organic molecules (Sec. 2a), we use a maximum depth of 3 in the $42d+30$ RACs, producing 156 potential descriptors, which reduce to 151 after discarding 5 descriptors that are constant (e.g., $^{lc}_{ax}I_0$ and $^{mc}_{all}T_0$) due to unchanged octahedral coordination in the data sets in this work (Supporting Information Tables S10). We add four variables (i.e., oxidation state, HF exchange and axial/equatorial ligand denticity) from our MCDL-25 set[51] to produce a final 155-variable set (RAC-155). The RAC-155 set is transferable to inorganic chemistry, with already good MCDL-25/KRR (Gaussian kernel) test set RMSE and MUE of 3.88 and 2.52 kcal/mol reduced to 1.80 and 1.00 kcal/mol with RAC-155 (Table 1). This performance is also superior to Coulomb matrix (CM)-based descriptors computed on high-spin geometries. Using either i) an L1 matrix difference kernel on sorted Coulomb matrices[69, 77] (CM-L1) or ii) eigenvalues[69] and a Laplacian kernel, as recommended in Ref.[72] (CM-ES), we obtain 10-30x higher RMSE and MUEs than for RAC-155 or MCDL-25 (Table 1).

**Table 1.** Test set KRR model prediction errors (RMSE and MUE) for spin-splitting energy (kcal/mol) for the Manhattan norm applied to sorted Coulomb matrices (CM-L1)[69, 77], the Coulomb matrix eigenspectrum representation with a Laplacian kernel (CM-ES)[72], our prior hybrid discrete-continuous descriptors (MCDL-25)[51] with a Gaussian kernel, and the full RAC-155 set introduced in this work with a Gaussian kernel.

| Feature set | RMSE (kcal/mol) | MUE (kcal/mol) |
|---|---|---|
| CM-L1 | 30.80 | 20.84 |
| CM-ES | 19.19 | 14.96 |
| MCDL-25 | 3.88 | 2.52 |
| RAC-155 | **1.80** | **1.00** |

Visualization with principal component analysis (PCA) of the key descriptor space dimensions with spin-splitting or molecular size variation overlaid reveals why CM-ES performs poorly in comparison to RACs (Figure 6). The first two principal components encode the majority of the feature space variation for both sets: 85% of CM-ES and 55% of RAC-155 (Supporting Information Figures S5 and S6). As expected[51, 70], CM-ES shows excessive molecule-size-dependent clustering that is not predictive of how metal electronic properties vary. As an example, homoleptic Fe(III) complexes with strong-field t-butylphenylisocyanide (pisc) and methylisocyanide (misc) ligands have comparable $\Delta E_{H-L}$ of 41 and 38 kcal/mol but differ in size substantially at 151 and 37 atoms, respectively (structures in Figure 6 inset). Despite comparable spin splitting, these molecules are on opposite ends of PC1 in the CM-ES PCA with no intermediate data (Figure 6). More broadly, no clustering is apparent in spin-splitting energies with CM-ES in comparison to the strong system size clustering (Figure 6).

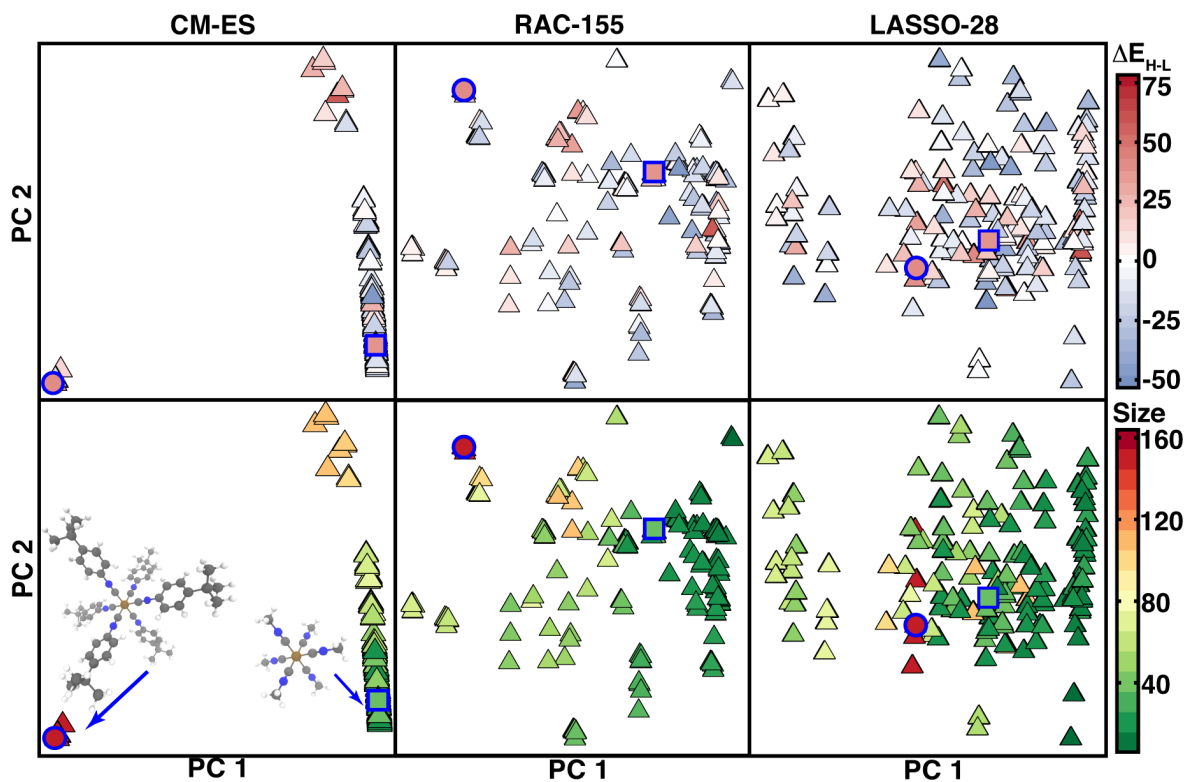

**Figure 6.** Projection of *spin-splitting* data set onto the first two principal components (arbitrary units) for the Coulomb matrix eigenspectrum (CM-ES, left), full revised AC set (RAC-155, center), and the LASSO-selected subset (LASSO-28, right). The PCA plots are colored by DFT-calculated spin splitting energy (top, scale bar in kcal/mol at right) and size (bottom, scale bar in number of atoms at right). Ball and stick structures of representative complexes $Fe(III)(pisc)_6$ and $Fe(III)(misc)_6$ (iron brown, nitrogen blue, carbon gray, and hydrogen white) are inset in the bottom left, and the associated data points are highlighted with a blue circle and square, respectively, in each plot.

In contrast, RAC-155 distributes data more evenly in the PCA with smaller size-dependence due to using both metal-centered and ligand-centered ACs in addition to truncating the depth of descriptors to three prior to feature selection (Figure 6). Improved RAC performance is also due to better representation of molecular similarity with apparent weak-field and strong-field groupings, assisting KRR learning[72] that relies on nearest neighbor influence for property prediction (Figure 6).

Spin splitting energies are well predicted by KRR with RAC-155, outperforming our previous MCDL-25 representation but at the initial cost of an order-of-magnitude increase (from 25 to 155) in feature space dimension. We thus apply feature selection techniques (Sec. 2c) to identify if AC subsets maintain predictive capability with smaller feature space size. Starting with Type 3 shrinkage methods we have previously employed[51], we carried out feature selection with an elastic net. Comparable CV scores were obtained for all α, and so we chose α = 1 (i.e., LASSO) (Supporting Information Figure S7). LASSO retained 28 features, eliminating over 80% of the features in RAC-155 with a 0.2 kcal/mol decrease in test RMSE and the best overall, sub-kcal/mol MUE (Table 2 and Supporting Information Table S11). PCA on LASSO-28 reveals even weaker size dependence than RAC-155 and closer pisc and misc species in PC space (Figure 6).

**Table 2.** Train and test set KRR model prediction errors (RMSE for train/test and MUE for test) for spin-splitting energy (in kcal/mol) for RAC-155 and down-selected subsets based on spin-splitting data using LASSO, univariate filters (UV), recursive feature elimination (RFE) based on MLR, and random forest (randF). The last results presented for comparison are the common feature subset (RAC-12), a proximal-only subset (PROX-23) of RAC-155, and the full RAC-155.

| Feature set | train        | test         |              |
|-------------|--------------|--------------|--------------|
|             | RMSE         | RMSE         | MUE          |
|             | (kcal/mol)   | (kcal/mol)   | (kcal/mol)   |
| LASSO-28    | 0.60         | **1.65**     | **0.96**     |
| UV-86       | 0.43         | 1.78         | 0.99         |
| RFE-43      | 0.41         | 2.50         | 1.20         |
| randF-41    | **0.40**     | 1.87         | 1.01         |
| randF-26    | 1.18         | 2.12         | 1.28         |
| RAC-12      | 1.31         | 2.90         | 1.86         |
| PROX-23     | 5.43         | 6.03         | 3.70         |
| RAC-155     | 0.55         | 1.80         | 1.00         |

Type 1 feature selection with UV filters (p <= 0.05) retains 86 features (UV-86, Supporting Information Table S12) and comparable performance to RAC-155, suggesting elimination of descriptors that have weak univariate correlation does not reduce KRR accuracy (Table 2 and Supporting Information Figure S8). Type 3 RFE with an embedded MLR model produces a flat CV error, with an absolute CV minimum at 43 retained features (i.e., RFE-43, Supporting Information Table S13 and Figure S9). The RFE-43 KRR model shows 0.5 kcal/mol and 0.2 kcal/mol worsened test RMSE and MUE, respectively, compared to RAC-155. Improved performance could possibly be obtained with a higher fidelity embedded model but at the cost of prohibitive computational time for feature selection (see Sec. 2c).

In addition to LASSO, we also employed the Type 3 random forest (randF) model, which has a suggested 1% MSE cutoff for feature selection, and by varying this cutoff we can vary feature set size. The standard 1% cutoff with random forest selects 41 features (randF-41), yielding KRR test RMSE/MUE within 0.1 kcal/mol of RAC-155 (Table 2 and Supporting Information Figure S10 and Table S14). We also truncate at 2% randF MSE to retain only 26 variables (randF-26), favorably reducing the feature space but slightly worsening test MUE relative to randF-41 or LASSO-28 by 0.2-0.3 kcal/mol, with other cutoffs yielding no KRR test error improvement (Table 2 and Supporting Information Tables S15-S16). In addition to average errors, error distributions are symmetric, and maximum errors track with RMSE/MUE: LASSO-28 yields the smallest (< 9 kcal/mol) maximum error (Supporting Information Figure S11).

The best-performing LASSO-28 set contains some features equivalent to those in MCDL-25: i) LASSO-28 $^{mc}_{all}\chi'_2$ and $^{mc}_{all}\chi'_3$ are similar to MCDL-25 $\Delta\chi$, ii) LASSO-28 $^{mc}_{all}S'_1$, $^{mc}_{all}Z_1$, and $^{lc}_{ax}\chi_1$ encode the size and identity of the ligand connecting atoms also present in MCDL-25, and iii) metal-identity, which was a discrete variable in MCDL-25, is represented by $^{mc}_{all}Z_0$ and $^{mc}_{all}\chi_0$ in LASSO-28. Our new difference-type RACs are well-represented (10 of 28 in LASSO-28), and only 5 of 28 are whole-ligand(4, e.g., $^{f}_{ax}I_3$) or whole-complex (1, $^{f}_{all}\chi_2$). Thus, *mc*, *lc*, and difference-derived RACs, all motivated by our prior observations of inorganic chemistry, are key to high accuracy predictions.

It is useful to understand the effect of feature selection method choice by identifying the number of common features among the three best-performing selected feature sets, LASSO-28, UV-86, and randF-41 (Figure 7). Only 12 features are common to the three subsets, which we designate RAC-12 (Supporting Information Table S17). In RAC-12, 7 of the retained descriptors are proximal, and 5 of 12 descriptors incorporate $\chi$ or $\Delta\chi$. All four of the retained distal properties in RAC-12 (e.g., $^{mc}_{all}\chi'_d$, $d=1,2,3$ and $^{mc}_{all}S'_1$) are of the newly introduced difference-derived AC type. A KRR model trained on RAC-12 produces test set RMSE and MUE 1.1 and 0.9 kcal/mol above the 13x larger RAC-155 but still significantly lower than the twice as large MCDL-25 (see Tables 1 and 2). Broadly, two thirds of all features are selected by at least one of the three best feature selection methods (Figure 7). Over 80% of the descriptors in randF-41 are also found in the larger UV-86, but fewer (31% of randF-41) are present in the smaller LASSO-28. Unique descriptors in randF-41 are *mc*-type, whereas unique LASSO-28 descriptors are non-local 2-depth or 3-depth standard ACs on ligands.

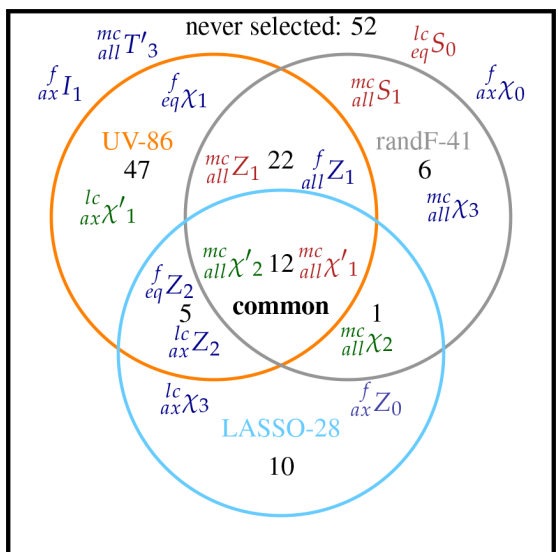

**Figure 7.** Venn diagram showing common descriptors among the three best performing subsets of RAC-155 returned by feature selection algorithms: UV-86, LASSO-28, and randF-41. A total of 12 common variables are found among all three sets, and other numbers refer to unique or common variables between sets. Example features are indicated, colored by classification (proximal in red, middle in green, and distal in blue).

We further classify the degree of locality in each feature set, as designated by the bond-wise path-length scales of information in the descriptors (i.e., proximal, middle, and distal, defined in Sec. 2b). We quantify the fraction of descriptors corresponding to each category in a feature set, e.g. the proximal fraction:

$$\text{frac}(proximal) = \frac{\text{num. of proximal RACs+2}}{\text{num. RACs+2}} \tag{10}$$

where the denominator only contains the RACs that can be assigned to proximal (the two ligand denticity variables are also included here), middle, or distal portions of the molecule, not oxidation state or HF exchange. Relative to RAC-155, all feature selection methods increase the proximal fraction, and we observe lowest MUEs in subsets with higher proximal fractions, i.e., over 0.3 in the best-performing LASSO-28 or in randF-41 and increased to nearly 0.5 when a higher MSE cutoff is used in random forest (i.e., randF-26, Figure 8). The higher-dimensional Type 1 UV-86 subset and Type 2 RFE-43 subset possess the most similar distributions to RAC-155 with still good performance likely due to relatively large feature set size (Figure 8). Modest feature space dimension (< 30) always gives higher proximal fraction than larger subsets.

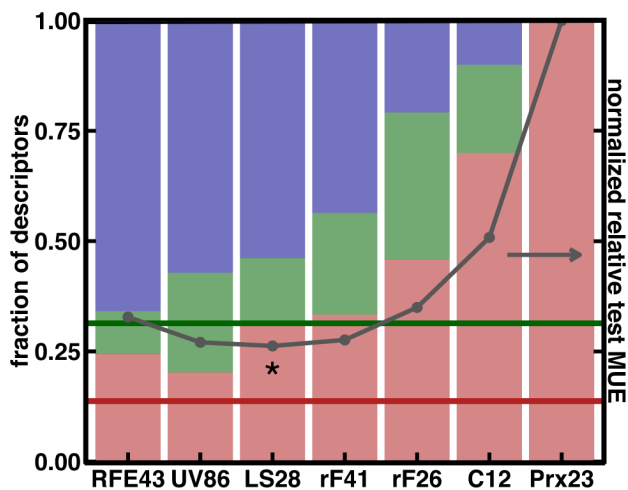

**Figure 8.** Fraction of selected descriptors that are proximal (red), middle (green) or distal (blue), as defined in the main text and depicted in Fig. 3 compared against RAC-155 reference fractions (dark red proximal fraction and green middle fraction as horizontal lines) along with their performance for spin-splitting prediction with KRR. The normalized relative test set spin-splitting MUE from a KRR model is shown in dark grey for each set, and the lowest test MUE is indicated with an asterisk. Sets are sorted left to right in decreasing distal fraction: RFE with MLR (RFE43); UV filter (UV86); LASSO (LS28), random forest with 1% (rF41) or 2% cutoff (rF26), common set (C12), and proximal-only (Prx23). HF exchange and oxidation state are not shown but are used in all models.

Given the high fraction of retained proximal descriptors in randF-26 and RAC-12, we also tested the suitability of a full proximal-only set of RACs and denticity variables along with oxidation state and HF exchange (PROX-23) for KRR model training (Supporting Information Table S18). This PROX-23 KRR model is the worst performing of all KRR models, including MCDL-25, with test RMSE and MUE of 6.0 and 3.7 kcal/mol, emphasizing the importance of beyond-proximal information present in both MCDL-25 and the feature sets selected in this work (Table 2). The superior performance of the LASSO-28 subset over the similarly-sized randF-26 also highlights the importance of second-shell and global descriptors, as 78% of the 18 features present in LASSO-28 that are absent from randF-26 are distal (e.g., $^{mc}_{all}\chi'_3$, $^{lc}_{eq}\chi'_2$, and $^{lc}_{ax}T_3$). Comparing randF-26 to the larger randF-41 set, which has a 0.3 kcal/mol lower test MUE, we observe that 12 of the 15 features present in randF-41 but omitted in randF-26 are distal.

### 4b. Descriptor Transferability to Bond Length Prediction.

A key advantage of our geometry-free RACs is that they enable bond length prediction[51] to facilitate accurate structure generation[63-64]. We first evaluate the predictive performance of our full AC set (RAC-155), the proximal subset (PROX-23), and spin-state splitting-selected feature sets (LASSO-28, randF-41, and randF-26) as well as the common subset (RAC-12) for training KRR models on minimum low-spin metal-ligand bond lengths (i.e., min($R_L$)) in the low-spin,

DFT geometry-optimized structures of complexes in the *spin-splitting* data set. If the complex is homoleptic and symmetric, there is only a single metal-ligand bond length in the low-spin complex that corresponds exactly to $\min(R_L)$, otherwise we take the minimum of the equatorial or axial metal-ligand bond length in order to predict a single property, as in previous work[51]. Except for PROX-23, all feature subsets yield RMSEs and MUEs around 1.4 and 0.5 and pm (i.e., 0.014 Å and 0.005 Å), respectively, with RAC-12 performing nearly as well (test RMSE: 1.6 pm, MUE: 0.6 pm) (Table 3). The overall best RMSE performance is observed for LASSO-28, better than for RAC-155, and all subsets have very slightly degraded (i.e., 0.05 pm worse) MUE performance compared to RAC-155 (Table 3). The PROX-23 set yields 2-3x larger errors (test RMSE: 2.7 pm and MUE: 1.8 pm), which is significantly worse than the smaller common set (RAC-12), indicating the critical importance of middle and distal features (Figure 9). Nevertheless, nearly all feature sets yield better prediction with a KRR model than our prior, proximally-weighted MCDL-25 set (neural network test RMSE: 2 pm).[51]

**Table 3.** Train and test set KRR model prediction errors (RMSE for train/test and MUE for test) for minimum low-spin bond length (in pm) for down-selected subsets of RAC-155 using LASSO and random forest (randF) on bond length data (denoted with suffix "B") shown first, as well as original spin-splitting feature sets (LASSO-28, randF-41, and randF-26), shown next. The randF-49B contains manually added HF exchange, which is excluded from automatically selected randF-48B. The last results presented for comparison are the common feature subset (RAC-12), a proximal-only subset (PROX-23) of RAC-155, and the full RAC-155.

| Feature set | train | test | |
| --- | --- | --- | --- |
| | RMSE (pm) | RMSE (pm) | MUE (pm) |
| LASSO-83B | 0.15 | 1.33 | **0.42** |
| randF-48B | 1.25 | 2.06 | 1.21 |
| randF-49B | 0.18 | 1.34 | 0.45 |
| LASSO-28 | **0.12** | **1.28** | 0.47 |
| randF-41 | 0.16 | 1.38 | 0.47 |
| randF-26 | 0.20 | 1.37 | 0.48 |
| RAC-12 | 0.16 | 1.62 | 0.59 |
| PROX-23 | 2.37 | 2.67 | 1.76 |
| RAC-155 | 0.16 | 1.33 | **0.42** |

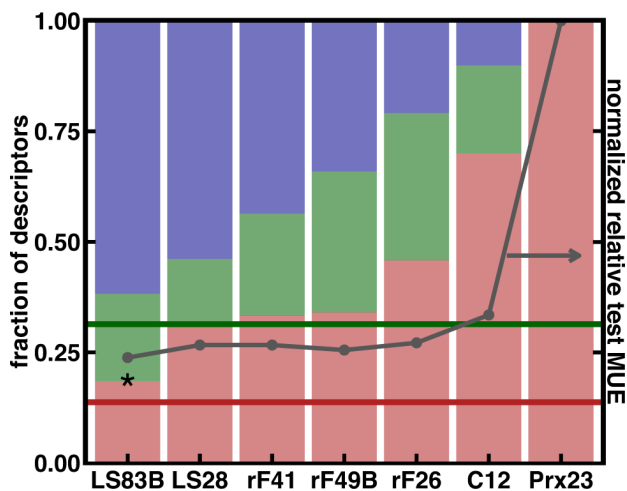

**Figure 9.** Fraction of selected descriptors that are proximal (red), middle (green) or distal (blue), as defined in the main text and depicted in Fig. 3 compared against RAC-155 reference fractions (dark red proximal fraction and green middle fraction as horizontal lines) along with their performance for low-spin bond length prediction with KRR. The normalized relative test set bond length MUE from a KRR model is shown in dark grey for each set, and the lowest test MUE is indicated with an asterisk. Sets are sorted left to right in decreasing distal fraction: LASSO on bond length (LS83B) or on spin-splitting data (LS28); random forest on spin-splitting (1%, rF41), on bond length data (1%, rF49B), higher cutoff on spin-splitting (or 2%, rF26); the spin-splitting-derived common set (C12); and proximal-only (Prx23). HF exchange and oxidation state are not shown but are used in all models.

We also carried out feature selection on the bond length data with LASSO and random forest to obtain new feature sets (denoted with a "B" suffix). With bond length data, LASSO and random forest retain larger feature sets of 83 and 48 RACs, respectively (LASSO-83B and randF-48B in Table 3, and Supporting Information Tables S19-S20 and Figures S12-S13). In KRR model training, LASSO-83B performs exactly the same as RAC-155 with half the features, whereas randF-48B has 2-3x larger errors (test RMSE: 2.1 pm, MUE: 1.2 pm). This degraded randF-48B performance occurs because HF exchange has been dropped at the 1% MSE random forest cutoff, producing a discontinuous jump in kernel hyperparameters (Table 3 and Supporting Information Table S4 and Figure S13). The indirect effect of HF exchange on bond length within a single complex is apparent[52], but across a wide data set of complexes, the role of HF exchange in bond length data is more easily missed by random forest than in the case of spin splitting. Manually adding HF exchange to the feature set (randF-49B) makes this set perform comparably in KRR model training to the other feature subsets (Table 3).

Comparison of random forest feature sets selected on bond lengths (randF-49B) and on spin splitting (randF-41) reveals differences in the underlying structure-property relationships. Both sets have 34 features in common, with an increased proximal fraction relative to RAC-155, but there is a slight bias toward middle features for the bond-length selected set (15 middle in randF-49B instead of 9 in randF-41) (Figure 9). The 15 unique features present in randF-49B but absent from randF-41 are weighted toward topological, size-derived effects with 5 $T$-type (e.g., $^{mc}_{all}T_1$), 2 I-type (e.g., $^{lc}_{ax}I_1$), and 4 $S$-type (e.g., $^{lc}_{eq}S_0$) RACs. Conversely, four of the seven features in randF-41 but absent from randF-49B are middle/distal and $\chi$-/Z-type (e.g., $^{mc}_{all}Z'_3$ and $^{lc}_{eq}\chi'_1$). Comparable KRR bond length prediction accuracy with both feature sets is due to similar data clustering: the ten nearest complexes to Fe(III)(pisc)$_6$ are largely unchanged between randF-49B and randF-41, but would differ substantially for RAC-12 and PROX-23 (Supporting Information Table S21). Thus evaluation of random forest feature set selection reveals structure-property-error relationships that may not be apparent from evaluating KRR model errors alone.

**4c. Descriptor Transferability to Redox Data.**

We now test the transferability of RAC descriptor sets to our *redox* data set for the prediction of M(II/III) gas phase ionization potentials (IPs) and aqueous redox potentials (see Figure 5). Here, all calculations are with B3LYP (20% exchange), and the oxidation state is no longer a fixed variable. Therefore, all feature sets have two fewer variables, but we retain the sets' original names. It might be expected that direct gas phase IPs are easier to learn than redox potentials, which incorporate composite and potentially opposing solvent and thermodynamic effects. However, we observe qualitatively similar KRR model performance and feature selection trends, and we thus summarize gas phase IP results briefly (Supporting Information Text S4, Tables S22-24, and Figures S14-S15). After removal of a single outlier molecule, RAC-155 yields test set RMSE and MUE values of 0.46 and 0.35 eV, respectively, or a 3% or 2% error relative to the 14.4 eV data set mean, and spin-splitting-selected subsets randF-41 or LASSO-28 produce the next lowest but slightly larger errors (Supporting Information Table S24 and Figures S16-S18).

Redox potentials (i.e., including thermodynamic and aqueous implicit solvent corrections) in the full *redox* data set range from 3.3 to 10.4 eV with a mean of 6.7 eV, and the gas phase IP outlier is not a redox potential outlier (Supporting Information Figure S17). The full RAC-155 set produces lower absolute errors with respect to gas phase IP (test: RMSE 0.40 eV, 6% error and MUE: 0.32 eV, 5% error) but higher relative errors due to the lower data set mean (Table 4). Feature selection on redox potentials from the *redox* data set retains 19 variables with LASSO (LASSO-19G), comparable to the size selected on gas phase IP but smaller than feature sets selected by LASSO on spin-splitting or bond length (Supporting Information Figure S19 and Table S25). LASSO-19G improves very slightly over RAC-155 (test RMSE: 0.38 eV and MUE: 0.31 eV), despite being 12% of the size of the full set (Table 4). Random forest on redox potential retains 38 features (randF-38G), improving over both LASSO-19G and RAC-155 (test RMSE: 0.31 eV, 5% error and MUE: 0.26 eV, 4% error) (Table 4 and Supporting Information Figure S20 and Table S26). Thus, comparable or reduced absolute errors and only slightly increased relative errors indicates that the combination of ionization potential, solvent, and thermodynamic corrections is only slightly more challenging to capture than IP alone.

**Table 4.** Train and test set KRR model prediction errors (RMSE for train/test and MUE for test) for redox potential (in eV) for down-selected subsets of RAC-155 using LASSO and random

forest (randF) on redox data (denoted with suffix "G") shown first, as well as original spin-splitting feature sets (LASSO-28, randF-41, and randF-26), shown next. The last results presented for comparison are the common feature subset (RAC-12) from all methods, a proximal-only subset (PROX-23) of RAC-155, and the full RAC-155.

| Feature set | train RMSE (eV) | test RMSE (eV) | test MUE (eV) |
|---|---|---|---|
| LASSO-19G | 0.17 | 0.38 | 0.31 |
| randF-38G | 0.16 | 0.33 | 0.26 |
| LASSO-28 | **0.10** | 0.46 | 0.35 |
| randF-41 | 0.32 | 0.31 | 0.26 |
| randF-26 | 0.35 | **0.29** | **0.23** |
| RAC-12 | 0.38 | 0.37 | 0.32 |
| PROX-23 | 0.87 | 0.91 | 0.78 |
| RAC-155 | 0.17 | 0.40 | 0.32 |

Evaluating the spin-splitting-selected feature subsets (LASSO-28, randF-41, and randF-26) and the common set (RAC-12) on the *redox* data set for redox potential prediction produces some of the lowest test errors of all sets (Table 4). The spin-splitting-selected randF-26 performs best (test RMSE: 0.29 eV, 4% error and MUE: 0.23 eV, 3% error), with the larger randF-41 performing nearly as well, whereas LASSO-28 has larger errors (e.g., test MUE of 0.35 eV) more comparable to RAC-155. The RAC-12 set exhibits its best relative performance for any property prediction so far (test RMSE: 0.37 eV and MUE: 0.32 eV), equivalent to the 13x larger full RAC-155 and substantially better than the proximal-only PROX-23 (test MUE: 0.78 eV, Table 4). The better performance of spin-splitting-selected sets on redox data could be due to i) the larger, more diverse data in the spin-splitting training set or ii) that our redox calculation implicitly requires knowledge of spin, as the redox potential is always evaluated from the ground state of the reduced species. However, separate prediction of high- or low-spin redox potentials yields similar accuracy, suggesting combined ground state and redox potential prediction does not increasing the difficulty of the learning task (Supporting Information Table S27).

Within the redox potential prediction subsets, a relationship between the prediction accuracy and fraction of descriptor type (i.e., proximal vs. distal) is less clear than for spin splitting or bond length (Figure 10). Simultaneously comparing locality and test set MUE across feature sets shows comparable performance for i) randF-38G with a proximal fraction below that of RAC-155, ii) the relatively high proximal and middle fractions in randF-26, and iii) and even relatively good performance in the RAC-12 minimal, proximal-heavy subset (Figure 10). Comparing the poorer performing spin-splitting-selected LASSO-28 to the redox-selected LASSO-19G reveals missing middle/distal *S*- or *I*-type RACs (e.g., $_{ax/eq}^{lc}I_3$, $_{ax/eq}^{lc}S'_1$) in the former.

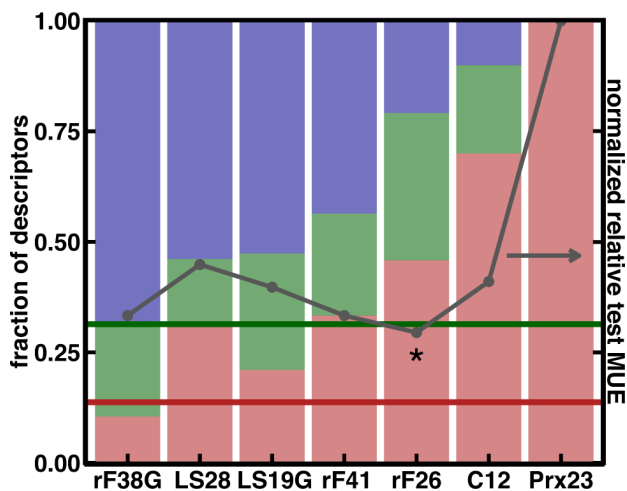

**Figure 10.** Fraction of selected descriptors that are proximal (red), middle (green) or distal (blue), as defined in the main text and depicted in Fig. 3 compared against RAC-155 reference fractions (dark red proximal fraction and green middle fraction as horizontal lines) along with their performance for redox potential prediction with KRR. The normalized relative test set redox potential MUE from a KRR model is shown in dark grey for each set, and the lowest test MUE is indicated with an asterisk. Sets are sorted left to right in decreasing distal fraction: random forest on redox potential (rF38G); LASSO on redox potential (LS19G) or spin-splitting (LS28); random forest on spin-splitting (1%, rF41 or 2%, rF26); spin-splitting common set (C12); and proximal-only (Prx23). HF exchange and oxidation state are not used in any models.

Examining descriptors in the better-performing, redox-selected randF-38G that are absent from similarly-sized spin-splitting-selected randF-41 reveals 10 *T*-type and 3 *I*-type RACs, seven *lc* 3-depth RACs, and two whole-ligand $_{eq}^{f}\chi_1$ and $_{eq}^{f}\chi_0$ RACs, indicating a preference for whole-complex-derived, and, in particular, connectivity information, consistent with observations of the importance of whole-ligand RACs in redox potentials[64]. Comparing instead the 17 common features in randF-38G and randF-41 reveals mostly *mc* RACs (e.g., $^{mc}_{all}Z_0$ and $^{mc}_{all}\chi'_1$,) similar to the metal and connecting atom information in MCDL-25[51].

### 4d. Overall Comparison of Best Descriptor Subsets.

Overall, Type 3 LASSO or random forest methods have provided the best price-performance trade-off for feature selection in KRR model training of transition metal complex properties on the data sets studied in this work. Although LASSO-28 produced the lowest KRR model test MUE of 0.96 kcal/mol, randF-41 (1% cutoff) and randF-26 (2% cutoff) produce similarly good 1.01-1.28 kcal/mol test MUEs on the *spin-splitting* data set and demonstrate somewhat better transferability to redox potential prediction on the *redox* data set. All three of these subsets are accurate for low-spin bond length prediction, with 1.3-1.4 pm test RMSE and 0.5 pm MUE that is only slightly worse relative to larger, bond-length-selected feature sets, randF-49B or LASSO-83B. The best redox potential prediction performance is achieved not with redox-selected randF-38G (test MUE: 0.26 eV), LASSO-19G (test MUE: 0.31 eV), or even the full RAC-155 (test MUE: 0.33 eV), but with the smaller spin-splitting selected randF-26 (test MUE: 0.23 eV). As an overall recommendation, we thus would select randF-26 for broad spin-splitting, bond length, and gas phase IP/redox potential prediction or LASSO-28 for only spin-splitting and bond length prediction.

To explore how feature space topology differs when using spin-splitting-selected features (randF-26 or randF-41) versus redox-selected features (randF-38G), we consider the example of Fe(II/III) complexes with triazolyl-pyridine ligands from the *redox* data set. In two cases, these homoleptic, bidentate complexes have a methyl group on the carbon adjacent to pyridinyl nitrogen (ligand 9, $E^0$ = 6.1 eV and ligand 23, $E^0$ = 6.0 eV), but in one case the methyl group is in the meta position with respect to the metal-coordinating pyridinyl nitrogen (ligand 8 with $E^0$ = 5.5 eV) (Supporting Information Figure S21). Ligands 8 and 9 contain a 1,2,3-triazole, whereas ligand 23 contains 1,2,4-triazole. Within randF-26 and randF-41, the high fraction of proximal or middle *mc* descriptors emphasizes differences between 1,2,3-triazole and 1,2,4-triazole rather than capturing the importance of the ligand-connecting atom adjacent methyl group. The additional distal *T*-, *I*- and *S*-type descriptors in randF-38G increase the relative importance of the metal-adjacent methyl groups over the order of ring substituents, correctly identifying the nearest neighbor of the ligand 9 complex as the ligand 23 complex (Supporting Information Text S5).

Although we have identified a feature set that is transferable across multiple properties when paired with a KRR model, there are still noteworthy differences in optimal feature sets obtained from random forest (i.e., spin-splitting randF-26/41, bond-length randF-49B, and redox randF-38G) that can inform our understanding of the degree of locality and nature of features needed for differing property prediction. To simplify this analysis, we classify $\chi$- and Z-derived RACs as electronic and *S*-, *I*-, and *T*- as topological (Figure 11). We confirm our earlier observations[51] of locality, especially in spin-splitting with randF-26/41: randF-49B and randF-38G both have more non-local (to the metal) and topological descriptors than randF-26/41.

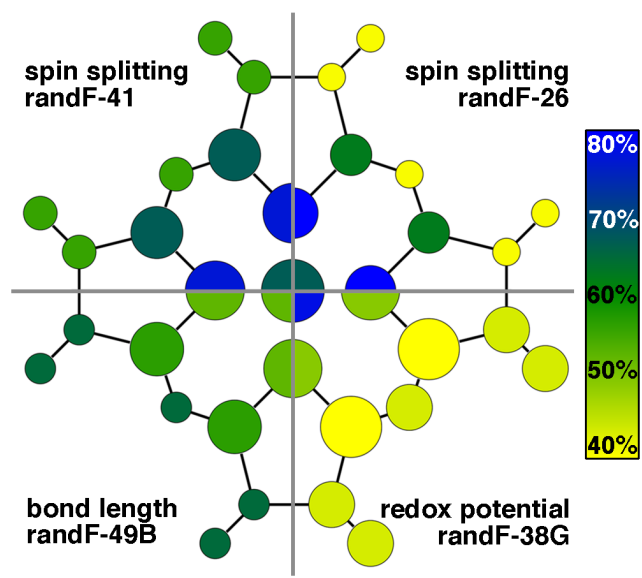

**Figure 11.** Schematic of relative proximity and electronic (blue) or topological (yellow) of feature sets on an iron-porphyrin complex. Feature sets are designated by their training data: spin splitting (randF-41 and randF-26, top), bond length (randF-49B, bottom left), and redox potential (randF-38G, bottom right). Atom sizes are scaled relative to the number of descriptor dimensions involving that atom (divided into first shell, second shell and other), scaled, with iron kept the same size in all sets. The color bar and absolute percentages of electronic and topological descriptors, as defined in the main text, is shown in inset right.

For direct ligand connection atoms, 80% of the descriptors are electronic for randF-26/41, but only 52% are electronic for randF-49B and 50% for randF-38G, which reflects the inclusion of additional first-shell *T*- and *I*-based RACs (Figure 11). Moving to the second shell shows increased topological fraction across all feature sets while preserving the first shell trends, with second shell descriptors around 65% electronic for the spin-splitting-selected randF-26/41 but only 40% electronic for randF-38G. LASSO-28 has an even stronger electronic, proximal bias than randF-26, possibly explaining its poorer performance for redox potential prediction (Supporting Information Figure S22). These observations suggest that overall ligand shape and size are more useful for prediction of redox potentials and bond lengths compared to spin splitting within the random forest model. These locality measures also highlight the features to be varied when collecting additional data in future work to enlarge the size of our *redox* data set and reach smaller ML prediction errors (e.g., 0.1 eV MUE) that would be beneficial for screening and discovery.

## 5. Conclusions

We have introduced a new series of revised autocorrelation (RACs) descriptors for machine learning of quantum chemical properties that extend prior ACs to incorporate modified starting points, scope over the molecule of interest, and incorporate differences of atomic properties. We first demonstrated superior performance of standard ACs on a large organic molecule test set, both showing the best yet performance for atomization energies based only on topological information, particularly when maximum topological distances were truncated at a modest maximum 3-bond distance.

We confirmed transferability of RACs from organic to inorganic chemistry with KRR model test set MUEs for the full RAC-155 set of 1 kcal/mol, in comparison to 15-20x larger errors from Coulomb-matrix-derived descriptors and 2-3x larger with our prior MCDL-25 set. We attribute this improvement to overestimation of size-dependence in CM descriptors and underestimation of distal effects in MCDL-25. LASSO or random forest feature selection yielded smaller subsets (LASSO-28 and randF-41, respectively) with improved or comparable sub- to 1-kcal/mol test MUEs. Restriction to a common set of descriptors identified by the three best feature selection tools yielded half as large spin-splitting errors (test MUE: 1.9 kcal/mol) compared to MCDL-25 with a still smaller 12 variable feature set. Both random forest as a feature selection tool and the spin-splitting-selected randF-26 showed the best combined transferability to bond length (0.005 Å test MUE) and redox potential (0.23 eV test MUE).

Random forest applied directly on bond length selected more topological features than for spin-splitting with equivalent locality bias. Selection based on redox potential data revealed redox potential to be both more non-local and more topological in nature than spin-splitting or bond lengths. However, invariant data-clustering within the trained KRR model means that no improvement in KRR test errors was observed with redox-selected features for redox potentials and only modest improvement using bond-length selected features for bond length prediction. Overall, this work provides both a prescription for machine learning models capable of making accurate predictions of inorganic complex quantum-mechanical properties and provides insight into locality in transition metal chemistry structure-property relationships.

# REFERENCES

1.      Curtarolo, S.; Hart, G. L.; Nardelli, M. B.; Mingo, N.; Sanvito, S.; Levy, O., The High-Throughput Highway to Computational Materials Design. *Nature materials* **2013,** *12*, 191-201.
2.      Greeley, J.; Jaramillo, T. F.; Bonde, J.; Chorkendorff, I.; Nørskov, J. K., Computational High-Throughput Screening of Electrocatalytic Materials for Hydrogen Evolution. *Nature materials* **2006,** *5*, 909-913.
3.      Nørskov, J. K.; Bligaard, T.; Rossmeisl, J.; Christensen, C. H., Towards the Computational Design of Solid Catalysts. *Nature chemistry* **2009,** *1*, 37-46.
4.      Jensen, P. B.; Bialy, A.; Blanchard, D.; Lysgaard, S.; Reumert, A. K.; Quaade, U. J.; Vegge, T., Accelerated DFT-Based Design of Materials for Ammonia Storage. *Chemistry of Materials* **2015,** *27*, 4552-4561.
5.      Hautier, G.; Fischer, C. C.; Jain, A.; Mueller, T.; Ceder, G., Finding Nature's Missing Ternary Oxide Compounds Using Machine Learning and Density Functional Theory. *Chemistry of Materials* **2010,** *22*, 3762-3767.
6.      Jain, A.; Hautier, G.; Moore, C. J.; Ong, S. P.; Fischer, C. C.; Mueller, T.; Persson, K. A.; Ceder, G., A High-Throughput Infrastructure for Density Functional Theory Calculations. *Computational Materials Science* **2011,** *50*, 2295-2310.
7.      Hautier, G.; Miglio, A.; Ceder, G.; Rignanese, G.-M.; Gonze, X., Identification and Design Principles of Low Hole Effective Mass P-Type Transparent Conducting Oxides. *Nature communications* **2013,** *4*, 2292.
8.      Bowman, D. N.; Bondarev, A.; Mukherjee, S.; Jakubikova, E., Tuning the Electronic Structure of Fe(II) Polypyridines via Donor Atom and Ligand Scaffold Modifications: A Computational Study. *Inorganic Chemistry* **2015,** *54*, 8786-8793.
9.      Eckert, H.; Bojorath, J., Molecular Similarity Analysis in Virtual Screening: Foundations, Limitations and Novel Approaches. *Drug Discov Today* **2007,** *12*, 225-233.
10.     Hachmann, J.; Olivares-Amaya, R.; Atahan-Evrenk, S.; Amador-Bedolla, C.; Sánchez-Carrera, R. S.; Gold-Parker, A.; Vogt, L.; Brockway, A. M.; Aspuru-Guzik, A., The Harvard

Clean Energy Project: Large-Scale Computational Screening and Design of Organic Photovoltaics on the World Community Grid. *The Journal of Physical Chemistry Letters* **2011,** *2*, 2241-2251.
11. Shu, Y.; Levine, B. G., Simulated Evolution of Fluorophores for Light Emitting Diodes. *The Journal of chemical physics* **2015,** *142*, 104104.
12. Virshup, A. M.; Contreras-García, J.; Wipf, P.; Yang, W.; Beratan, D. N., Stochastic Voyages into Uncharted Chemical Space Produce a Representative Library of All Possible Drug-Like Compounds. *Journal of the American Chemical Society* **2013,** *135*, 7296-7303.
13. Kirkpatrick, P.; Ellis, C., Chemical Space. *Nature* **2004,** *432*, 823-823.
14. Meredig, B.; Agrawal, A.; Kirklin, S.; Saal, J. E.; Doak, J. W.; Thompson, A.; Zhang, K.; Choudhary, A.; Wolverton, C., Combinatorial Screening for New Materials in Unconstrained Composition Space with Machine Learning. *Physical Review B* **2014,** *89*, 094104.
15. Behler, J., Perspective: Machine Learning Potentials for Atomistic Simulations. *The Journal of Chemical Physics* **2016,** *145*, 170901.
16. Behler, J., Representing Potential Energy Surfaces by High-Dimensional Neural Network Potentials. *Journal of Physics: Condensed Matter* **2014,** *26*, 183001.
17. Lorenz, S.; Groß, A.; Scheffler, M., Representing High-Dimensional Potential-Energy Surfaces for Reactions at Surfaces by Neural Networks. *Chemical Physics Letters* **2004,** *395*, 210-215.
18. Artrith, N.; Morawietz, T.; Behler, J., High-Dimensional Neural-Network Potentials for Multicomponent Systems: Applications to Zinc Oxide. *Physical Review B* **2011,** *83*, 153101.
19. Behler, J.; Parrinello, M., Generalized Neural-Network Representation of High-Dimensional Potential-Energy Surfaces. *Physical Review Letters* **2007,** *98*, 146401.
20. Prudente, F. V.; Neto, J. J. S., The Fitting of Potential Energy Surfaces Using Neural Networks. Application to the Study of the Photodissociation Processes. *Chemical Physics Letters* **1998,** *287*, 585-589.
21. Mones, L.; Bernstein, N.; Csanyi, G., Exploration, Sampling, and Reconstruction of Free Energy Surfaces with Gaussian Process Regression. *Journal of Chemical Theory and Computation* **2016,** *12*, 5100-5110.
22. Smith, J. S.; Isayev, O.; Roitberg, A. E., Ani-1: An Extensible Neural Network Potential with DFT Accuracy at Force Field Computational Cost. *Chemical Science* **2017**.
23. Snyder, J. C.; Rupp, M.; Hansen, K.; Muller, K.-R.; Burke, K., Finding Density Functionals with Machine Learning. *Physical Review Letters* **2012,** *108*, 253002.
24. Su, X.; Kulik, H. J.; Jamison, T. F.; Hatton, T. A., Anion-Selective Redox Electrodes: Electrochemically Mediated Separation with Heterogeneous Organometallic Interfaces. *Advanced Functional Materials* **2016,** *26*, 3394-3404.
25. Mills, K.; Spanner, M.; Tamblyn, I., Deep Learning and the Schr\" Odinger Equation. *arXiv preprint arXiv:1702.01361* **2017**.
26. Yao, K.; Parkhill, J., Kinetic Energy of Hydrocarbons as a Function of Electron Density and Convolutional Neural Networks. *Journal of Chemical Theory and Computation* **2016,** *12*, 1139-1147.
27. Snyder, J. C.; Rupp, M.; Hansen, K.; Blooston, L.; Müller, K.-R.; Burke, K., Orbital-Free Bond Breaking via Machine Learning. *The Journal of Chemical Physics* **2013,** *139*, 224104.
28. Yao, K.; Herr, J. E.; Parkhill, J., The Many-Body Expansion Combined with Neural Networks. *The Journal of Chemical Physics* **2017,** *146*, 014106.

29. Hase, F.; Valleau, S.; Pyzer-Knapp, E.; Aspuru-Guzik, A., Machine Learning Exciton Dynamics. *Chemical Science* **2016,** *7*, 5139-5147.
30. Li, Z.; Kermode, J. R.; De Vita, A., Molecular Dynamics with on-the-Fly Machine Learning of Quantum-Mechanical Forces. *Physical Review Letters* **2015,** *114*, 096405.
31. Botu, V.; Ramprasad, R., Adaptive Machine Learning Framework to Accelerate Ab Initio Molecular Dynamics. *International Journal of Quantum Chemistry* **2015,** *115*, 1074-1083.
32. Pilania, G.; Gubernatis, J.; Lookman, T., Multi-Fidelity Machine Learning Models for Accurate Bandgap Predictions of Solids. *Computational Materials Science* **2017,** *129*, 156-163.
33. Pilania, G.; Mannodi-Kanakkithodi, A.; Uberuaga, B. P.; Ramprasad, R.; Gubernatis, J. E.; Lookman, T., Machine Learning Bandgaps of Double Perovskites. *Scientific Reports* **2016,** *6*, 19375.
34. Gomez-Bombarelli, R.; Aguilera-Iparraguirre, J.; Hirzel, T. D.; Duvenaud, D.; Maclaurin, D.; Blood-Forsythe, M. A.; Chae, H. S.; Einzinger, M.; Ha, D. G.; Wu, T.; Markopoulos, G.; Jeon, S.; Kang, H.; Miyazaki, H.; Numata, M.; Kim, S.; Huang, W.; Hong, S. I.; Baldo, M.; Adams, R. P.; Aspuru-Guzik, A., Design of Efficient Molecular Organic Light-Emitting Diodes by a High-Throughput Virtual Screening and Experimental Approach. *Nature Materials* **2016,** *15*, 1120-1127.
35. Pyzer-Knapp, E. O.; Li, K.; Aspuru-Guzik, A., Learning from the Harvard Clean Energy Project: The Use of Neural Networks to Accelerate Materials Discovery. *Advanced Functional Materials* **2015,** *25*, 6495-6502.
36. Ma, X.; Li, Z.; Achenie, L. E. K.; Xin, H., Machine-Learning-Augmented Chemisorption Model for CO2 Electroreduction Catalyst Screening. *The Journal of Physical Chemistry Letters* **2015,** *6*, 3528-3533.
37. Mannodi-Kanakkithodi, A.; Pilania, G.; Huan, T. D.; Lookman, T.; Ramprasad, R., Machine Learning Strategy for Accelerated Design of Polymer Dielectrics. *Scientific Reports* **2016,** *6*, 20952.
38. Huan, T. D.; Mannodi-Kanakkithodi, A.; Ramprasad, R., Accelerated Materials Property Predictions and Design Using Motif-Based Fingerprints. *Physical Review B* **2015,** *92*, 014106.
39. Pilania, G.; Wang, C.; Jiang, X.; Rajasekaran, S.; Ramprasad, R., Accelerating Materials Property Predictions Using Machine Learning. *Scientific Reports* **2013,** *3*, 2810.
40. Lee, J.; Seko, A.; Shitara, K.; Tanaka, I., Prediction Model of Band-Gap for AX Binary Compounds by Combination of Density Functional Theory Calculations and Machine Learning Techniques. *Physical Review B* **2016,** *93*, 115104.
41. Bartók, A. P.; Kondor, R.; Csányi, G., On Representing Chemical Environments. *Physical Review B* **2013,** *87*, 184115.
42. Ghiringhelli, L. M.; Vybiral, J.; Levchenko, S. V.; Draxl, C.; Scheffler, M., Big Data of Materials Science: Critical Role of the Descriptor. *Physical Review Letters* **2015,** *114*, 105503.
43. Huang, B.; von Lilienfeld, O. A., Communication: Understanding Molecular Representations in Machine Learning: The Role of Uniqueness and Target Similarity. *The Journal of Chemical Physics* **2016,** *145*, 161102.
44. Rupp, M.; Tkatchenko, A.; Müller, K.-R.; von Lilienfeld, O. A., Fast and Accurate Modeling of Molecular Atomization Energies with Machine Learning. *Physical Review Letters* **2012,** *108*, 058301.
45. De, S.; Bartok, A. P.; Csanyi, G.; Ceriotti, M., Comparing Molecules and Solids across Structural and Alchemical Space. *Physical Chemistry Chemical Physics* **2016,** *18*, 13754-13769.

46. Maggiora, G.; Vogt, M.; Stumpfe, D.; Bajorath, J., Molecular Similarity in Medicinal Chemistry: Miniperspective. *Journal of medicinal chemistry* **2013,** *57*, 3186-3204.
47. Wang, J.; Wolf, R. M.; Caldwell, J. W.; Kollman, P. A.; Case, D. A., Development and Testing of a General Amber Force Field. *Journal of computational chemistry* **2004,** *25*, 1157-1174.
48. Kubinyi, H., QSAR and 3D QSAR in Drug Design. Part 1: Methodology. *Drug Discov Today* **1997,** *2*, 457-467.
49. Benson, S. W.; Cruickshank, F.; Golden, D.; Haugen, G. R.; O'neal, H.; Rodgers, A.; Shaw, R.; Walsh, R., Additivity Rules for the Estimation of Thermochemical Properties. *Chemical Reviews* **1969,** *69*, 279-324.
50. Schütt, K. T.; Glawe, H.; Brockherde, F.; Sanna, A.; Müller, K. R.; Gross, E. K. U., How to Represent Crystal Structures for Machine Learning: Towards Fast Prediction of Electronic Properties. *Physical Review B* **2014,** *89*, 205118.
51. Janet, J. P.; Kulik, H. J., Predicting Electronic Structure Properties of Transition Metal Complexes with Neural Networks. *Chemical Science* **2017,** *8*, 5137-5152.
52. Ioannidis, E. I.; Kulik, H. J., Towards Quantifying the Role of Exact Exchange in Predictions of Transition Metal Complex Properties. *Journal of Chemical Physics* **2015,** *143*, 034104.
53. Ashley, D. C.; Jakubikova, E., Ironing out the Photochemical and Spin-Crossover Behavior of Fe (II) Coordination Compounds with Computational Chemistry. *Coordination Chemistry Reviews* **2017**.
54. Bowman, D. N.; Jakubikova, E., Low-Spin Versus High-Spin Ground State in Pseudo-Octahedral Iron Complexes. *Inorganic Chemistry* **2012,** *51*, 6011-6019.
55. Gani, T. Z.; Kulik, H. J., Where Does the Density Localize? Convergent Behavior for Global Hybrids, Range Separation, and DFT+ U. *Journal of chemical theory and computation* **2016,** *12*, 5931-5945.
56. Ioannidis, E. I.; Kulik, H. J., Ligand-Field-Dependent Behavior of Meta-GGA Exchange in Transition-Metal Complex Spin-State Ordering. *Journal of Physical Chemistry A* **2017,** *121*, 874-884.
57. Deeth, R. J., The Ligand Field Molecular Mechanics Model and the Stereoelectronic Effects of d and S Electrons. *Coordination Chemistry Reviews* **2001,** *212*, 11-34.
58. Shriver, D. F.; Atkins, P. W., *Inorganic Chemistry*. 3rd ed.; W.H. Freeman and Co.: 1999.
59. Halcrow, M. A., Structure: Function Relationships in Molecular Spin-Crossover Complexes. *Chemical Society Reviews* **2011,** *40*, 4119-4142.
60. Létard, J.-F.; Guionneau, P.; Goux-Capes, L., Towards Spin Crossover Applications. In *Spin Crossover in Transition Metal Compounds III*, Springer: 2004; pp 221-249.
61. Bignozzi, C. A.; Argazzi, R.; Boaretto, R.; Busatto, E.; Carli, S.; Ronconi, F.; Caramori, S., The Role of Transition Metal Complexes in Dye-Sensitized Solar Devices. *Coordination Chemistry Reviews* **2013,** *257*, 1472-1492.
62. Harvey, J. N.; Poli, R.; Smith, K. M., Understanding the Reactivity of Transition Metal Complexes Involving Multiple Spin States. *Coordination chemistry reviews* **2003,** *238*, 347-361.
63. Ioannidis, E. I.; Gani, T. Z. H.; Kulik, H. J., Molsimplify: A Toolkit for Automating Discovery in Inorganic Chemistry. *Journal of Computational Chemistry* **2016,** *37*, 2106-2117.

64. Janet, J. P.; Gani, T. Z. H.; Steeves, A. H.; Ioannidis, E. I.; Kulik, H. J., Leveraging Cheminformatics Strategies for Inorganic Discovery: Application to Redox Potential Design. *Industrial & Engineering Chemistry Research* **2017,** *56*, 4898-4910.
65. Broto, P.; Moreau, G.; Vandycke, C., Molecular Structures: Perception, Autocorrelation Descriptor and Sar Studies: System of Atomic Contributions for the Calculation of the N-Octanol/Water Partition Coefficients. *European Journal of Medicinal Chemistry* **1984,** *19*, 71-78.
66. Devillers, J.; Domine, D.; Guillon, C.; Bintein, S.; Karcher, W., Prediction of Partition Coefficients (Log P Oct) Using Autocorrelation Descriptors. *SAR and QSAR in Environmental Research* **1997,** *7*, 151-172.
67. Broto, P.; Devillers, J., *Autocorrelation of Properties Distributed on Molecular Graphs*. Kluwer Academic Publishers: Dordrecht, The Netherlands: 1990.
68. Puzyn, T.; Leszczynski, J.; Cronin, M. T., *Recent Advances in QSAR Studies: Methods and Applications*. Springer Science & Business Media: 2010; Vol. 8.
69. Montavon, G.; Hansen, K.; Fazli, S.; Rupp, M. In *Learning Invariant Representations of Molecules for Atomization Energy Prediction*, Advances in Neural Information Processing Systems, Pereira, F.; Burges, C. J. C.; Bottou, L.; Weinberger, K. Q., Eds. Curran Associates, Inc.: 2012; pp 440-448.
70. Collins, C. R.; Gordon, G. J.; Anatole von Lilienfeld, O.; Yaron, D. J., Constant Size Molecular Descriptors for Use with Machine Learning. *arXiv preprint arxiv:1701.06649* **2017**.
71. Hansen, K.; Biegler, F.; Ramakrishnan, R.; Pronobis, W., Machine Learning Predictions of Molecular Properties: Accurate Many-Body Potentials and Nonlocality in Chemical Space. *Journal of Physical Chemistry Letters* **2015,** *6*, 2326-2331.
72. Hansen, K.; Montavon, G.; Biegler, F.; Fazli, S.; Rupp, M.; Scheffler, M.; von Lilienfeld, O. A.; Tkatchenko, A.; Müller, K.-R., Assessment and Validation of Machine Learning Methods for Predicting Molecular Atomization Energies. *Journal of Chemical Theory and Computation* **2013,** *9*, 3404-3419.
73. Ramakrishnan, R.; Dral, P. O.; Rupp, M.; von Lilienfeld, O. A., Quantum Chemistry Structures and Properties of 134 Kilo Molecules. **2014,** *1*, 140022.
74. Faber, F. A.; Hutchison, L.; Huang, B.; Gilmer, J.; Schoenholz, S. S.; Dahl, G. E.; Vinyals, O.; Kearnes, S.; Riley, P. F.; Anatole von Lilienfeld, O., Machine Learning Prediction Errors Better Than DFT Accuracy. *ArXiv e-prints* **2017,** *1702*, arXiv:1702.05532.
75. Hastie, T.; Tibshirani, R.; Friedman, J., *The Elements of Statistical Learning*. Springer New York: 2009; Vol. 18, p 764.
76. Huo, H.; Rupp, M., Unified Representation for Machine Learning of Molecules and Crystals. *arXiv:1704.06439v2* **2017,** *1704*.
77. Ramakrishnan, R.; Dral, P. O.; Rupp, M.; von Lilienfeld, O. A., Big Data Meets Quantum Chemistry Approximations: The Delta-Machine Learning Approach. *Journal of Chemical Theory and Computation* **2015,** *11*, 2087-96.
78. Kier, L. B., A Shape Index from Molecular Graphs. *Quantitative Structure-Activity Relationships* **1985,** *4*, 109-116.
79. Gani, T. Z. H.; Ioannidis, E. I.; Kulik, H. J., Computational Discovery of Hydrogen Bond Design Rules for Electrochemical Ion Separation. *Chemistry of Materials* **2016,** *28*, 6207-6218.
80. Guyon, I.; Elisseeff, A., An Introduction to Variable and Feature Selection. *Journal of Machine Learning Research* **2003,** *3*, 1157-1182.

81. Saeys, Y.; Inza, I.; Larrañaga, P., A Review of Feature Selection Techniques in Bioinformatics. *Bioinformatics* **2007,** *23*, 2507-2517.
82. Guyon, I. E., André An Introduction to Variable and Feature Selection. *Journal of Machine Learning Research* **2003,** *3*, 1157-1182.
83. Tibshirani, R., Regression Shrinkage and Selection via the Lasso. *Journal of the Royal Statistical Society. Series B (Methodological)* **1996,** *58*, 267-288.
84. Zou, H.; Hastie, T., Regularization and Variable Selection via the Elastic Net. *Journal of the Royal Statistical Society: Series B (Statistical Methodology)* **2005,** *67*, 301-320.
85. Breiman, L., Random Forests. *Machine Learning* **2001,** *45*, 5-32.
86. Genuer, R.; Poggi, J.-M.; Tuleau-Malot, C., Variable Selection Using Random Forests. *Pattern Recognition Letters* **2010,** *31*, 2225-2236.
87. R Core Team, R: A Language and Environment for Statistical Computing. Vienna, Austria, 2015.
88. Karatzoglou, A.; Smola, A.; Hornik, K.; Zeileis, A., Kernlab - an S4 Package for Kernel Methods in R. *Journal of Statistical Software* **2004,** *11*, 1-20.
89. Krueger, T.; Braun, M., Cvst: Fast Cross-Validation via Sequential Testing. 2013.
90. Friedman, J.; Hastie, T.; Tibshirani, R., Regularization Paths for Generalized Linear Models via Coordinate Descent. *Journal of statistical software* **2010,** *33*, 1-22.
91. Wing, J.; Kuhn, M., Caret: Classification and Regression Training. 2017.
92. Liaw, A.; Wiener, M., Classification and Regression by Randomforest. *R News* **2002,** *2*, 18-22.
93. Kramida, A., Ralchenko, Yu., Reader, J. and NIST ASD Team NIST Atomic Spectra Database (Version 5.3). http://physics.nist.gov/asd (accessed March 14, 2017).
94. Konezny, S. J.; Doherty, M. D.; Luca, O. R.; Crabtree, R. H.; Soloveichik, G. L.; Batista, V. S., Reduction of Systematic Uncertainty in DFT Redox Potentials of Transition-Metal Complexes. *The Journal of Physical Chemistry C* **2012,** *116*, 6349-6356.
95. Roy, L. E.; Jakubikova, E.; Guthrie, M. G.; Batista, E. R., Calculation of One-Electron Redox Potentials Revisited. Is It Possible to Calculate Accurate Potentials with Density Functional Methods? *The Journal of Physical Chemistry A* **2009,** *113*, 6745-6750.
96. Baik, M.-H.; Friesner, R. A., Computing Redox Potentials in Solution: Density Functional Theory as a Tool for Rational Design of Redox Agents. *The Journal of Physical Chemistry A* **2002,** *106*, 7407-7412.
97. Stephens, P. J.; Devlin, F. J.; Chabalowski, C. F.; Frisch, M. J., Ab Initio Calculation of Vibrational Absorption and Circular Dichroism Spectra Using Density Functional Force Fields. *Journal of Physical Chemistry* **1994,** *98*, 11623-11627.
98. Becke, A. D., Density-Functional Thermochemistry. III. The Role of Exact Exchange. *Journal of Chemical Physics* **1993,** *98*, 5648-5652.
99. Lee, C.; Yang, W.; Parr, R. G., Development of the Colle-Salvetti Correlation-Energy Formula into a Functional of the Electron Density. *Physical Review B* **1988,** *37*, 785--789.
100. Hay, P. J.; Wadt, W. R., Ab Initio Effective Core Potentials for Molecular Calculations. Potentials for the Transition Metal Atoms Sc to Hg. *Journal of Chemical Physics* **1985,** *82*, 270-283.
101. Kästner, J.; Carr, J. M.; Keal, T. W.; Thiel, W.; Wander, A.; Sherwood, P., DL-FIND: An Open-Source Geometry Optimizer for Atomistic Simulations. *The Journal of Physical Chemistry A* **2009,** *113*, 11856-11865.

102. Wang, L.-P.; Song, C., Geometry Optimization Made Simple with Translation and Rotation Coordinates. *The Journal of Chemical Physics* **2016,** *144*, 214108.
103. Petachem. http://www.petachem.com. (accessed February 23, 2017).
104. Ufimtsev, I. S.; Martinez, T. J., Quantum Chemistry on Graphical Processing Units. 3. Analytical Energy Gradients, Geometry Optimization, and First Principles Molecular Dynamics. *Journal of Chemical Theory and Computation* **2009,** *5*, 2619-2628.
105. Saunders, V. R.; Hillier, I. H., A "Level–Shifting" Method for Converging Closed Shell Hartree–Fock Wave Functions. *International Journal of Quantum Chemistry* **1973,** *7*, 699-705.
106. Klamt, A.; Schuurmann, G., Cosmo: A New Approach to Dielectric Screening in Solvents with Explicit Expressions for the Screening Energy and Its Gradient. *Journal of the Chemical Society, Perkin Transactions 2* **1993,** *2*, 799-805.
107. Liu, F.; Luehr, N.; Kulik, H. J.; Martínez, T. J., Quantum Chemistry for Solvated Molecules on Graphical Processing Units Using Polarizable Continuum Models. *Journal of Chemical Theory and Computation* **2015,** *11*, 3131-3144.
108. Bondi, A., Van Der Waals Volumes and Radii. *The Journal of physical chemistry* **1964,** *68*, 441-451.